\documentclass[aps,twocolumn]{revtex4}
\usepackage{epsfig}
\usepackage{amsmath}

\begin{document}

\title
{Heat transport along a chain of coupled quantum harmonic oscillators}

\author{M\'ario J. de Oliveira}

\affiliation{
Instituto de F\'{\i}sica,
Universidade de S\~{a}o Paulo, \\
Rua do Mat\~ao, 1371,
05508-090 S\~{a}o Paulo, S\~{a}o Paulo, Brazil}

\date{\today}

\begin{abstract}

We study the heat transport properties of a chain of coupled quantum harmonic
oscillators in contact at its ends with two heat reservoirs
at distinct temperatures. Our approach is based on the use of
an evolution equation for the density operator which is a canonical
quantization of the classical Fokker-Planck-Kramers equation.
We set up the evolution equation for the covariances and obtain the
stationary covariances at the stationary states
from which we determine the thermal conductance in closed form
when the interparticle interaction is small.
The conductance is finite in the thermodynamic limit
implying an infinite thermal conductivity.


\end{abstract}

\maketitle


Fifty years ago, Rieder, Lebowitz, and Lieb \cite{rieder1967}
introduced and exactly solved a microscopic model for thermal conduction,
that consisted of a chain of coupled classic harmonic oscillators
with its ends in contact with heat reservoirs at distinct temperatures.
Using this model they provided a rigorous proof of
the well known result that the thermal
conductance, the ratio between the heat current and the 
temperature difference, is finite regardless of the chain length
\cite{peierls1955}.
This result amounts to say that Fourier's law does not hold because
the conductivity, which is the product of the conductance and the chain length,
becomes infinite when the length increases without bounds.
The reason for the occurrence of a finite conductance is that
the excitations in ordered systems with harmonic interactions
travel balistically.
To get the Fourier's law, new ingredients should be added to
the harmonic model in order to transform the ballistic into
a diffusive motion. Such ingredients include
anharmonic potentials 
\cite{lepri1997,aoki2001,eckmann2004,mai2007,bernardin2011,roy2012},
self-consistent reservoirs
\cite{bolsterli1970,bonetto2004,pereira2004,delfini2006},
energy conserving noise
\cite{dhar2011,landi2013,landi2014},
and others
\cite{narayan2002,lepri2003,casati2003,basile2009,iacobucci2010}.

In the model studied by Rieder, Lebowitz, and Lieb \cite{rieder1967},
the oscillators were under the action of conservative forces
except the first and the last which in addition were subject to
dissipating-fluctuating forces representing the
contact with heat reservoirs. They are composed by
a dissipative force, proportional to the velocity, and a 
Gaussian white noise with zero mean and 
variance proportional to the temperature. 
The equations of motion are understood as Langevin equations, 
and the equation governing the time evolution of the probability
density is a Fokker-Planck-Kramers (FPK) equation
\cite{kampen1981,gardiner1983,risken1984,tome2015book}.

Here, we study a quantum version of
the model studied by Rieder, Lebowit, and Lieb \cite{rieder1967}. 
We have exactly calculated the thermal conductance
in the regime of small interparticle interaction
and reached a similar result that the conductance
is finite regardless of the length of the chain.
However, as should be expected the conductance
is not independent of temperature, as is the case 
of the classical version.
It vanishes in the limit of zero temperature and
saturates at the classical value at high temperatures.
Our approach is based on a quantum version of the FPK
equation coming from a canonical quantization of the ordinary
FPK equation, recently introduced \cite{oliveira2016}, 
and differs from other approaches regarding the treatment of
quantum dissipation
\cite{zurcher1990,saito2000,dhar2003,segal2003,dhar2006,roy2008,rivas2010,
dhar2012,asadian2013,freitas2014,nicacio2015}.
These approaches include the use of quantum Langevin
equations \cite{zurcher1990,dhar2006,roy2008}, and the use of the rotating
wave approximation and Lindblad master equation
to describe the contact with heat reservoirs \cite{rivas2010,asadian2013}.
These approaches as well as ours, when applied to the harmonic
chain, predict a finite conductance regardless of the chain length
\cite{zurcher1990,asadian2013}.
It worth mentioning that the approach we use leads to a proper
thermalization in the sense that the Gibbs equilibrium state
is the stationary solution of the quantum FPK equation
when the reservoirs have the same temperatures \cite{oliveira2016}.


The model we consider is a chain of $L$ particles of equal masses
interacting through
a harmonic potential. The quantum hamiltonian of the system is given by
\begin{equation}
{\cal H} = \frac{1}{2m}\sum_{i=1}^L p_i^2 + \frac{k_0}2 \sum_{i=1}^L q_i^2
- k_1 \sum_{i=1}^{L-1} q_i q_{i+1},
\end{equation}
where $m$ is the mass of the particles, $k_0$ is the spring constant
and $k_1$ is the interparticle interacting parameter.
The position $q_i$ and momentum $p_i$
obey the usual commutation relation, $[q_i,p_j]=i\hbar\delta_{ij}$.
To describe the contact of the system with heat reservoirs we use 
the quantum Fokker-Planck-Kramers (FPK) equation \cite{oliveira2016}
with the first particle of the chain in contact
with a heat reservoir A at temperature $T_A$ and the last 
with a reservoir B at temperature $T_B$. The quantum
FPK equation reads \cite{oliveira2016}
\begin{equation}
i\hbar\frac{\partial{\rho}}{\partial t} = [{\cal H},\rho] - [q_1,J_1] - [q_L,J_L],
\label{2}
\end{equation}
where $\rho$ is the density matrix, and $J_1$ and $J_2$ are given by
\begin{equation}
J_i = -\frac{\gamma}2(\rho g_i + g_i^\dagger\rho)
- \frac{\gamma m}{i\hbar\beta_i}[q_i,\rho],
\end{equation}
where $\gamma$ is the dissipation parameter and
$\beta_i=1/k_BT_i$, with $T_1=T_A$ and $T_L=T_B$.
The operator $g_i$ is given by
\begin{equation}
g_i = - \frac{ m}{i\hbar\beta_i}
(e^{\beta_i{\cal H}}q_i e^{-\beta_i{\cal H}} - q_i).
\end{equation}
When the temperatures are the same, the Gibbs density
$\rho_0=(1/Z)e^{-\beta{\cal H}}$ is the stationary equilibrium
solution of the quantum FPK equation (\ref{2}) because it makes
each one of the $J_i$ to vanish and $[{\cal H},\rho_0]=0$.

The operator $g_i$ can be written in a form involving nested commutators
\begin{equation}
g_i = p_i + \frac{\beta_i}{2!}[{\cal H},p_i]
+ \frac{\beta_i^2}{3!}[{\cal H},[{\cal H},p_i]]
+ \frac{\beta_i^3}{4!}[{\cal H},[{\cal H},[{\cal H},p_i]]] + \ldots
\label{15}
\end{equation}
Taking into account the bilinear form of ${\cal H}$ and that the coordinates
are not coupled to the momenta, we notice that the odd terms in this expansion
are linear combinations of the momenta only, and that the even terms
are linear combinations of the coordinates only. From these properties,
it follows that $g_i$ is a linear combination of
the positions and momenta,
\begin{equation}
g_i =\sum_{j=1}^L (a_{ij}p_j + i b_{ij}q_j),
\end{equation}
where the coefficients $a_{ij}$ and $b_{ij}$ depend on the
temperature $T_i$ and on the parameters of the Hamiltonian.
In addition, using the fact that the odd terms are Hermitian
and that the even terms are anti-Hermitian it follows that
the coefficient of $p_j$ is real and the coefficient of $q_j$
is pure imaginary so that $a_{ij}$ and $b_{ij}$ are real.


Next, we wish to write down evolution equations for 
the averages of quantities of interest in our analysis. 
The evolution equation for a certain average 
$\langle f\rangle={\rm Tr}\{f\rho\}$ of an operator $f$ can
be obtained from the quantum FPK equation (\ref{2}) and it is given by
\begin{equation}
i\hbar\frac{d}{dt}\langle f\rangle = \langle[f,H]\rangle
- {\rm Tr}[f,q_1]J_1 - {\rm Tr}[f,q_L]J_L,
\label{27}
\end{equation}
\begin{equation}
{\rm Tr}[f,q_i]J_i = 
- \frac{\gamma}2 \langle g_i[f,q_i] \rangle
- \frac{\gamma}2 \langle[f,q_i]g_i^\dagger\rangle
- \frac{\gamma m}{i\hbar\beta_i} \langle[[f,q_i],q_i]\rangle. 
\end{equation}
From this formula we get at once
\begin{equation}
\frac{d}{dt}\langle {\cal H}\rangle = \Phi_1 + \Phi_L
\end{equation}
where $\Phi_i=(1/m){\rm Tr}p_i J_i$ 
is the energy flux from reservoir $i$ to the chain.

Using formula (\ref{27}) we get the evolution equation for the correlations
among the coordinates and momenta, $x_{ij}=\langle q_i q_j\rangle$,
$y_{ij}=\langle p_i p_j\rangle$, and $z_{ij}=\langle q_i p_j\rangle$, 
\begin{equation}
\frac{d}{dt} x_{ij}= \frac1m (z_{ij}+\tilde{z}_{ij}),
\label{15a}
\end{equation}
\begin{equation}
\frac{d}{dt} z_{ij} = - \sum_{k=1}^L x_{ik} G_{kj} + \frac{1}{m} y_{ij}
- \frac{\gamma}2 \sum_{k=1}^L a_{jk}(\tilde{z}_{ki} + z_{ik}),
\label{15b}
\end{equation}
\begin{multline}
\frac{d}{dt}y_{ij}
= - \sum_{k=1}^L (G_{ik} z_{kj} + \tilde{z}_{ik} G_{kj})
+ 2\gamma m \delta_{ij}
\left(\frac{\delta_{j1}}{\beta_1} + \frac{\delta_{jL}}{\beta_L}\right) \\
- \gamma \sum_{k=1}^L (a_{jk} y_{ki} + a_{ik} y_{kj})
+ \frac{\gamma\hbar}2 (b_{ji} + b_{ij}),
\label{15c}
\end{multline}
where $\tilde{z}_{ij}=\langle p_i q_j\rangle$, and
$G_{ij}$ are the elements of a tridiagonal matrix $G$, with
$G_{ii} = k_0$, and $G_{i,i+1}=G_{i+1,i}=-k_1$.
In these equations, we are setting the coefficients $a_{ij}$ and $b_{ij}$ to
vanish unless $i=1$ or $i=L$. We wee that equations (\ref{15a}), 
(\ref{15b}), and (\ref{15c}) make up a closed set of equations for
the correlations $x_{ij}$, $y_{ij}$, and $z_{ij}$ by recognizing that
$\tilde{z}_{ji}=z_{ij} -i\hbar\delta_{ij}$.

In the stationary state, $\Phi_L=-\Phi_1$ and $\Phi=\Phi_L$
may thus be understood as the heat flux from reservoir B to
reservoir A through the chain.
Using equations (\ref{15a}) and (\ref{15c}),
we can show that $z_{21}=z_{32}=\ldots=z_{L,L-1}$, that $z_{ji}=-z_{ij}$,
and that $\Phi=(k_1/m)z_{i+1,i}$, a relation that will be used
to determine the conductance. 

To simplify the evolution equations, we will subtract the 
equilibrium solution, which we denote by $x_{ij}^{\rm e}$,
$y_{ij}^{\rm e}$, and $z_{ij}^{\rm e}$. By equilibrium solution we
mean the stationary solution of (\ref{15a}), (\ref{15b}),
and (\ref{15c}) for the case when both temperatures
of the reservoirs are the same and equal to $T=(T_1+T_L)/2$.
In equilibrium $z_{ij}^{\rm e}=\tilde{z}_{ij}^{\rm e}=0$ if $i\neq j$
and $z_{ii}^{\rm e}=-\tilde{z}_{ii}^{\rm e}=i\hbar/2$. The correlations
$x_{ij}^{\rm e}$ and $y_{ij}^{\rm e}$ are given by
\begin{equation}
\sum_{k=1}^L x_{ik}^{\rm e} G_{kj} = \frac{1}{m} y_{ij}^{\rm e},
\end{equation}
\begin{equation}
\sum_{k=1}^L (A_{jk} y_{ki}^{\rm e} + A_{ik} y_{kj}^{\rm e})
= \frac{\hbar}2 (B_{ji} + B_{ij}) 
+ \frac{2 m}{\beta} \delta_{ij}(\delta_{j1}+\delta_{jL}),
\end{equation}
where $\beta=1/k_BT$, and $A_{ik}$ and $B_{ik}$
are the values of $a_{ij}$ and $b_{ij}$ that one obtains
by replacing both $\beta_1$ and $\beta_L$ by $\beta$.  
The quantities $A_{ij}$ and $B_{ij}$ are nonzero only when $i=1$
or $i=L$ in which case they hold the property
$A_{Lj}=A_{1,L+1-j}$ and $B_{Lj}=B_{1,L+1-j}$.

We define the deviations $X_{ij}\Delta T=x_{ij}-x_{ij}^{\rm e}$,
$Y_{ij}\Delta T= y_{ij}-y_{ij}^{\rm e}$, $Z_{ij}\Delta T=z_{ij}-z_{ij}^{\rm e}$,
and $\tilde{Z}_{ij}\Delta T=\tilde{z}_{ij}-\tilde{z}_{ij}^{\rm e}$
from the equilibrium solution, where $\Delta T=T_L-T_1$.
We remark that $\tilde{Z}_{ij}=Z_{ji}$.
We wish, in the following, to write down evolution equations for
the variables $X_{ij}$, $Y_{ij}$, and $Z_{ij}$ for small values of
$\Delta T$. In this regime the evolution equation reads
\begin{equation}
\frac{d}{dt} X_{ij}= \frac1m (Z_{ij}+\tilde{Z}_{ij}),
\label{18a}
\end{equation}
\begin{equation}
\frac{d}{dt} Z_{ij} = - \sum_{k=1}^L X_{ik} G_{kj} + \frac{1}{m} Y_{ij}
- \frac{\gamma}2 \sum_{k=1}^L A_{jk}(\tilde{Z}_{ki} + Z_{ik}),
\label{18b}
\end{equation}
\begin{multline}
\frac{d}{dt}Y_{ij}
= - \sum_{k=1}^L (G_{ik} Z_{kj} + \tilde{Z}_{ik} G_{kj})
+ \gamma m k_B \delta_{ij} (-\delta_{j1} + \delta_{jL}) \\
- \gamma \sum_{k=1}^L(A_{jk} Y_{ki} + A_{ik} Y_{kj})
-\gamma (C_{ji}+C_{ij}),
\label{18c}
\end{multline}
where
\begin{equation}
C_{ij} \Delta T= 
\sum_{k=1}^L (a_{ik}-A_{ik}) y_{kj}^{\rm e}
- \frac{\hbar}2 (b_{ij}-B_{ij}),
\label{39}
\end{equation}
The quantities $C_{ij}$ are nonzero only when $i=1$ or $i=L$
in which case they hold the property $C_{Lj}=-C_{1,L+1-j}$.

These equations are written in matrix form as
\begin{equation}
\frac{d}{dt} X= \frac1m (Z+Z^\dagger),
\label{19a}
\end{equation}
\begin{equation}
\frac{d}{dt} Z = - X G + \frac{1}{m} Y - \gamma  Z A^\dagger,
\label{19b}
\end{equation}
\begin{equation}
\frac{d}{dt}Y
= - (G Z + Z^\dagger G) +\gamma D - ( Y A^\dagger+ AY) -\gamma (C^\dagger+C),
\label{19c}
\end{equation}
where the matrix $D$ has only two nonzero elements, which are
$D_{11} = - mk_B$ and $D_{LL} = mk_B$. 
It should be noted that all the entries of $X$, $Y$, and $Z$ are real
and that $X$ and $Y$ are symmetric matrices. All entries of
matrices $G$, $A$ and $C$ are also real and $G$ is symmetric.
The matrices $A$ and $C$ have nonzero values only on the first
and the last row.

In the stationary state we are left with the equations
\begin{equation}
Z = - Z^\dagger,
\label{17a}
\end{equation}
\begin{equation}
Y = m(X G  + \gamma Z A^\dagger),
\label{17b}
\end{equation}
\begin{equation}
Y = m(G X - \gamma A Z),
\label{17c}
\end{equation}
\begin{equation}
\gamma D -\gamma (Y A^\dagger + A Y) - \gamma(C + C^\dagger) = G Z - Z G.
\label{17d}
\end{equation}
In the classical limit, $C$ vanishes and $A$ becomes a matrix whose
only nonzero entries are $A_{11}=1$ and $A_{LL}=1$, and we
recover the equations obtained by Rieder, Lebowitz, and Lieb \cite{rieder1967}.


Our next step is to seek the solution of equations (\ref{17a}), (\ref{17b}),
(\ref{17c}), and (\ref{17d}). To this end we follow the reasoning 
put forward by Rieder, Lebowitz, and Lieb \cite{rieder1967}.
We start by observing that the left-hand side of equation (\ref{17d})
is a bordered matrix, that is, a matrix whose nonvanishing entries
are found only in the first and last rows and columns. 
Therefore, the right-hand side $GZ-ZG$ should also be bordered.
Using the definition of $G$ and the relation $Z_{ji} = - Z_{ij}$,
that comes from (\ref{17a}) and the property that $Z_{ij}$ is real,
it follows that $Z$ is an antisymmetric Toeplitz matrix, that is,
a matrix of the type
\begin{equation}
Z_{ij} = \left\{
\begin{array}{ll}
\;\;\varphi_{j-i},  & \;\; i<j, \\
\;\; 0,      & \;\; i=j, \\
-\varphi_{i-j},     & \;\; i>j.
\end{array}
\right.
\label{24}
\end{equation}

From equations (\ref{17b}) and (\ref{17c}), we get the following
relation between $X$ and $Z$, 
\begin{equation}
X G - G X  = - \gamma (A Z + Z A^\dagger).
\label{25}
\end{equation}
The right-hand side of this equation is again a bordered matrix
and so is the matrix $XG-GX$. Since $X$ is required to be symmetric,
a solution for $X$ is an antisymmetric Hankel matrix, that is, 
a matrix of the following form 
\begin{equation}
X_{ij} = \left\{
\begin{array}{ll}
\;\;\psi_{i+j-1},  & \;\; i+j<L+1, \\
\;\; 0,      & \;\; i+j=L+1, \\
-\psi_{2L+1-i-j},     & \;\; i+j>L+1.
\end{array}
\right.
\label{30}
\end{equation}
Replacing (\ref{30}) into (\ref{25}), we find $\psi_i$ in terms
of $\varphi_i$
\begin{equation}
\frac{k_1}{\gamma} \psi_\ell =
- \sum_{j=1}^L \eta_{\ell,j-1} A_j \varphi_{|\ell-j+1|},
\label{31}
\end{equation}
$1\leq \ell< L$, where $A_j$ stands for $A_{1j}$ and
$\eta_{\ell,j}=-1,0,1$ according to whether
$j<\ell$, $j=\ell$, $j>\ell$, respectively.

From equations (\ref{17b}) and (\ref{17c}), it is straitforward to show
that $Y$ is also an antisymmetric Hankel matrix, that is,
a matrix of the following form 
\begin{equation}
Y_{ij} = \left\{
\begin{array}{ll}
\;\;\theta_{i+j-1},  & \;\; i+j<L+1, \\
\;\; 0,      & \;\; i+j=L+1, \\
-\theta_{2L+1-i-j},     & \;\; i+j>L+1,
\end{array}
\right.
\label{33}
\end{equation}
and that $\theta_\ell$ is related to $\psi_\ell$ by
\begin{equation}
\theta_\ell = m (k_0 \,\psi_\ell - k_1 \psi_{\ell+1} - k_1 \psi_{\ell-1}),
\label{34} 
\end{equation}
$1\leq \ell< L$, where $\psi_0=0$ and $\psi_L=0$.

Since $\psi_\ell$ is related to $\varphi_\ell$ by expression (\ref{31}),
then equation (\ref{34}) gives $\theta_\ell$ in terms of $\varphi_\ell$.
To get a closed equation for $\varphi_\ell$ we use equation (\ref{17d})
to obtain another relation between $\varphi_\ell$ and $\theta_\ell$,
namely
\begin{equation}
\frac{k_1}{\gamma} \varphi_\ell  = -\frac{m}2  k_B \delta_{\ell 1}
- \sum_{j=1}^L \eta_{\ell,L+1-j} A_j \theta_{L-|L+1-\ell-j|} - C_{\ell},
\label{35a}
\end{equation}
$1\leq \ell< L$, where $C_{\ell}$ stands for $C_{1\ell}$. 
Therefore, equations (\ref{31}), (\ref{34}), and (\ref{35a})
constitute the desired closed equations for the variables $\varphi_\ell$.
To solve them we need to know $A_\ell$ and $B_\ell$ because $C_\ell$ 
is related to these quantities by 
\begin{equation}
C_\ell = - \frac12 \sum_{k=1}^L \frac{dA_k}{dT}\, y_{k\ell}^{\rm e}
+ \frac{\hbar}4 \frac{dB_\ell}{dT},
\label{39a}
\end{equation}
which follows from (\ref{39}), where $B_\ell$ stands for $B_{1\ell}$.
We recall that $A_\ell$ and $B_\ell$ depend on $T$ and are the
values of $a_{1\ell}$ and $b_{1\ell}$ obtaining by setting $T_1$
equal to $T$. 

Explicit solutions of equations (\ref{31}), (\ref{34}), and (\ref{35a}) 
are very cumbersome to find, but closed solutions can be found
when the interparticle interaction is small.
Thus, from now on we will confine ourselves to the case where
the interparticle interacting parameter $k_1$ is small.
To this end, we first notice that
the quantities $A_\ell$ and $B_\ell$ are of the order $k_1^{\ell-1}$
and so is $C_\ell$, a result that follows from expansion (\ref{15}).
Thus, from equations   
(\ref{31}), (\ref{34}), and (\ref{35a}) it follows that
$\varphi_\ell$, $\theta_\ell$, and $\psi_\ell$ are of order greater or
equal $k_1$, except $\theta_1$ and $\psi_1$, which are
\begin{equation}
\theta_1 = -\frac{1}{2A_1}(m k_B  + 2C_1),
\qquad\qquad
\psi_1 = \frac{1}{m k_0} \theta_1
\end{equation}
Using equation (\ref{31}), we see that $\varphi_1$ is of the order $k_1$
and given by
\begin{equation}
\varphi_1 = \frac{k_1}{\gamma A_1} \psi_1.
\end{equation}


The conductance $K$ is defined as the ratio $K=\Phi/\Delta T$
in the limit $\Delta T\to0$.
To determine $K$, we recall that,
in the stationary state, the heat flux 
$\Phi=(k_1/m)z_{i+1,i}=(k_1/m)Z_{i+1,i}\Delta T$
so that the conductance is 
$K=(k_1/m)Z_{i+1,i}=-(k_1/m)\varphi_1$,
leading us to following expression for the conductance
\begin{equation}
K = \frac{k_1^2}{2m^2 k_0\gamma A_1^2} (m k_B  + 2C_1).
\end{equation}
In this formula the value of $C_1$ is
\begin{equation}
C_1 = - \frac12 \frac{dA_1}{dT}\, y_{11}^{\rm e}
+ \frac{\hbar}4 \frac{dB_1}{dT},
\end{equation}
Using formula (\ref{15}), we obtain explicit expression for
$A_1$ and $B_1$, which for $k_1=0$ reads
\begin{equation}
A_1 = \frac{\sinh\beta\hbar\omega}{\beta\hbar\omega},
\qquad\qquad
B_1 = \frac{m(\cosh\beta\hbar\omega - 1)}{\beta\hbar} ,
\end{equation}
where $\omega=\sqrt{k_0/m}$.
When $k_1=0$, $y_{11}^{\rm e}=mE$ where
\begin{equation}
E = \hbar\omega\left(\frac{1}{e^{\beta\hbar\omega}-1}+\frac12\right).
\end{equation}

A straightforward algebra leads us to the result
\begin{equation}
K =  \frac{K_0}{A_1 k_B}\frac{dE}{dT},
\qquad\qquad
K_0 = k_B\frac{k_1^2}{2m\gamma k_0}.
\end{equation}
In the classical limit, $E=k_BT$ and $A_1=1$ so that $K=K_0$.
Thus $K_0$ is the classical conductance,
a result obtained by Rieder, Lebowitz, and Lieb \cite{rieder1967},
in the regime of small $k_1$. An explicit form for $K$ is
\begin{equation}
K = \frac{K_0(\beta\hbar\omega)^3}
{2\sinh\beta\hbar\omega(\cosh\beta\hbar\omega-1)}.
\end{equation}

As regards the behavior with temperature, 
this result is qualitatively similar to the ones obtained
by other approaches \cite{roy2008,asadian2013,nicacio2015}.
At high temperatures all these approaches
give the classical conductance obtained in referece
\cite{rieder1967}, and vanish when $T\to0$. However, the behavior
at low temperature is distinct. Our results give the
behavior $\beta^3 e^{-2\beta\hbar\omega}$ for the conductance
whereas the reference \cite{roy2008}, for instance, gives the behavior
$\beta^{1/2} e^{-\beta\hbar\omega}$.


In conclusion, we have used a quantum FPK equation to cond the
contact of a chain of coupled quantum harmonic oscillators
with heat reservoirs at distinct temperatures.
Starting from the quantum FPK equation, 
we have set up evolution equations for the covariances
and solved them in the stationary regime to get the
thermal conductance. An exact closed form for the conductance
was obtained for small values of the interparticle interacting parameter.
The conductance was found to be finite regardless of the chain length,
implying an infinite conductivity and thus the absence of Fourier's law.
This is a consequence of the ballistic motion of
phonons that occurs in a system with harmonic interactions 
where the phonons do not interact.

The quantum FPK equation we use holds two important properties
with relevant consequences to our analysis. In equilibrium, that is,
when the heat baths have the same temperatures, its stationary state
is the Gibbs state. Second, it is a canonical quantization
of the ordinary FPK equation \cite{oliveira2016} and,
as a consequence it turns into this equation,
in the classical limit. 
Thus, in the classical limit our approach becomes identical to
that of Rieder, Lebowitz, and Lieb \cite{rieder1967}.
We remark that our quantum approach differs from other
approaches used to study the thermal transport such as 
the use of quantum Langevin equations 
with dissipation proportional do velocity
or the use of Lindblad dissipators.
Although the quantum FPK equation we used here has not been
derived from a full quantum system that includes the reservoirs,
we think that the canonical quantization might give support
for the validity of the quantum FPK equation, although this
procedure does not give an unambiguous prescription for
obtaining a quantum version of a classical equation
\cite{oliveira2016}. The present calculation of the conductance,
on the other hand, might give indirectly the desired support if
we uderstand that the conductance obtained here is a 
reasonable result.


\end{document}